\documentstyle[prl,aps,multicol]{revtex}
\renewcommand{\narrowtext}{\begin{multicols}{2} \global\columnwidth20.5pc}
  \renewcommand{\widetext}{\end{multicols} \global\columnwidth42.5pc}
\begin{document}
\title{Waveguide diffusion modes and slowdown of D'yakonov-Perel'
spin relaxation in narrow 2-D semiconductor channels.}
\author{A. G. Mal'shukov}
\address{Institute of Spectroscopy, Russian Academy of Sciences, 142092
Troitsk, Moscow Region, Russia}
\author{K. A. Chao}
\address{Department of Theoretical Physics, Lund University, S-223 62
Lund, Sweden}
\maketitle

\begin{abstract}
We have shown that in narrow 2D semiconductor channels the D'yakonov-Perel'
spin relaxation rate is strongly reduced. This relaxation slowdown appears
in special waveguide diffusion modes which determine the propagation of spin
density in long channels. Experiments are suggested to detect the
theoretically predicted effects. A possible application is a field effect
transistor operated with injected spin current.
\end{abstract}



\narrowtext
In connection to the possible use of electron spin for storage and
information transfer in quantum computers~\cite{divi}, there has been much
recent studies on spin transport in semiconductor nanostructures. Among
various materials for the so-called {\it spintronic devices}, a favorite
candidate is III-V semiconductors because their spin-orbit split
conduction bands have unusual electron spin dynamics. The spin-orbit
interaction (SOI) in such materials has the form $H_{so}$={\bf h}({\bf k})$%
\cdot ${\bf s}, where {\bf s} is the electron spin, and the direction and
magnitude of the vector {\bf h}({\bf k)} depend on the electron momentum
{\bf k}. When an electron propagates, its spin precesses around the
direction of {\bf h}({\bf k}). In materials with narrow band gap, the
magnitude $|${\bf h}({\bf k})$|$, and hence the angle of spin rotation can
be varied by simply applying a gate voltage~\cite{nitta}. While this effect
is valuable to those devices which make use ofballistically propagating
spin-polarized electrons~\cite{datta}, it requires a high degree of coherency
in electron propagation, because scattering from an impurity or the boundary
changes the direction of {\bf h}({\bf k}). Scatterings thus randomize spin
precession. If the mean scattering time $ \tau $ is shorter than the
precession frequency $|${\bf h}({\bf k})$|^{-1}$, the spin dynamics is more
like a diffusive relaxation in the angular space with the D'yakonov-Perel'
relaxation rate of the order of
1/$\tau _{s}$=$\frac{1}{2}\tau h^{2}$({\bf k})~\cite{dp}. However, the
evolution of an inhomogeneous spin polarization at a spatial region is
determined not only by the diffusive randomization of the local spin
orientation, but also by a balance of the incoming and outgoung spin
currents. These spin currents depend on the polarization gradient, and
contain a component associated to the spin precession. Hence, it is
impossible to determine unambigously which part of the spin relaxation is
due to the diffusion of spin orientation in the angular space, and which is
due to the spin diffusion in the coordinate space. On the other hand, one
can examine the evolution of a given spin polarization by studying the
dynamics of individual eigenmodes of the spin diffusion equation. After a
sufficiently long time, only those modes with lowest spin relaxation rate
survive, and this rate will be characteristic to the spin relaxation of the
system under consideration.

In this paper we will study the spin diffusion in a 2D channel made from
a quantum well with growth direction along the $y$ axis. We choose our
coordinate system to have $x$ axis along the channel, and the boundaries of
the channel are marked at $z$=$\pm$$d/2$. The channel width $d$ is much
shorter than the spin precession length $L_{s}$=$v_{_{F}}/|h({\bf k}_{F})|$,
where $v_{_{F}}$ is the electron Fermi velocity. On the other hand, the
width $d$ is much longer than the Fermi wavelength, so that the electron
motion perpendicular to the channel (along $z$ axis) is semiclassical. We
will show that the long-time evolution of spin polarization in such a
channel is dominated by a waveguide {\it diffusion} mode. The spin
relaxation time of this mode is slowed down dramatically by a factor
$L_s^2/d^2$ with respect to the D'yakonov-Perel' relaxation time $\tau_s$,
which is the typical spin relaxation time for bulk materials and quantum
wells. Furthermore, the diffusion mode produces a periodic rotation of the
spin polarization from point to point along the 2D channel. Such a
phenomenon is similar to the spin precession of ballistic electrons
considered in Ref.~\onlinecite{datta}, but with a difference that in our
case the rotating spin has the spin quantum number 1 instead of 1/2. This
difference is due to the fact that our nonequilibrium spin density is
represented by two-particle excitations with electrons above the Fermi level
and holes below it. As an example, we will demonstrate the oscillations in
the resistance of a diffusive channel when the spin polarization is injected
and probed via ferromagnets at both ends of the channel. We will also
discuss qualitatively the effects of spin relaxation slowdown in a channel
on the weak localization behavior of transport parameters.

In terms of the creation and destruction operators $c_{{\bf k}}^{\dag}(t)$
and $c_{{\bf k}}(t)$, the spin density {\bf S}({\bf r},$t$) of a 2D
electron gas is defined as
\[
{\bf S}({\bf r,}t)=\sum_{{\bf k,q}}e^{i{\bf qr}}\langle Tr[c_{{\bf k+q}%
}^{\dag }(t){\bf s}c_{{\bf k}}(t)]\rangle \, ,
\]
where the average is taken over the ground state of the electron system.
Taking into account the SOI and the less important Coulomb effects on spin
density excitations, from the quasiclassical kinetic equation for
$<$$Tr[c_{{\bf k+q}}^{\dag}(t){\bf s}c_{{\bf k}}(t)]$$>$, or by means of the
standard perturbation theory~\cite{us2,ilp,pp}, one can derive the diffusion
equation for the spin density {\bf S}({\bf r},$t$).

It is convenient to represent {\bf S}({\bf r},$t$) in the basis set of three
eigenstates of the $z$-component $J_z$ of the angular momentum operator
{\bf J} which has the angular momentum quantum number $J$=1. Accordingly, we
introduce $\psi_1$=$(S_x$-$iS_y)/\sqrt{2}$, $\psi_0$=$S_z$, and
$\psi_{-1}$=-$(S_x$+$iS_y)/\sqrt{2}$, where the indices 1, 0, and -1 are the
three eigenvalues of $J_z$. In term of this basis set, the diffusion
equation is expressed as
\begin{equation}  \label{de}
\frac{\partial\psi}{\partial t} + \tau\langle (-i{\bf v}_{_F}\cdot {\bf %
\nabla}_{{\bf r}} + {\bf h}_{{\bf k}}\cdot {\bf J})^2\rangle_{dir}\psi = I(%
{\bf r},t) \, ,
\end{equation}
where $\langle ...\rangle _{dir}$ is an angular average over the Fermi line
and $I({\bf r},t)$ represents a possible source of spin oriented electrons
inside the channel. The corresponding eigenmode equation is simply
\begin{equation} \label{deA}
\tau\langle (-i{\bf v}_{_F}\cdot{\bf \nabla}_{{\bf r}} +
{\bf h}_{{\bf k}}\cdot {\bf J})^2\rangle_{dir}\psi = \Gamma\psi \, .
\end{equation}
The eigenvalue $\Gamma$ is equal to the relaxation rate of the corresponding
diffusion eigenmode. Eq.~(\ref{de}) is valid if
$v_{_F}|{\bf \nabla}_{{\bf r}}\psi|$$\ll$$1/\tau$ and
$|h(k_F)|$$\ll$$1/\tau$. The second inequality can not be satisfied for some
high mobility InAs based quantum wells with a strong SOI, because in which
$v_{_F}/|h(k_F)|$$\leq$500~nm according to Ref.~\onlinecite{nitta} and
\onlinecite{datta}. Eq.~(\ref{de}) can be generalized to the region
$|h(k_F)|$$\geq$$1/\tau$, a situation outside the scope of the present work.

In a semiconductor quantum well, there are two contributions to the SOI
{\bf h}({\bf k)}: the Dresselhaus term has its origin in the bulk crystal
spin splitting~\cite{dress}, and the Rashba term is due to the asymmetric
potential profile in the quantum well~\cite{rash}. The Rashba term is the
dominating one in quantum wells made of narrow gap semiconductors such as
InAs~\cite{nitta}. Therefore, for the systems of our interest, we will
retain only the Rashba term of SOI, which can be expressed as
$h_{z}$=$\alpha k_{x}$, $h_{x}$=-$\alpha k_{y}$, and $h_{y}$=0. After taking
the average over the direction of ${\bf k}_{F}$, Eq.~(\ref{deA}) becomes
\begin{equation}
D(i\frac{\partial }{\partial x}+m^{*}\alpha J_{z})^{2}\psi +D(i\frac{%
\partial }{\partial z}-m^{*}\alpha J_{x})^{2}\psi =\Gamma \psi \,,
\label{de2}
\end{equation}
where $D$$\equiv $$v_{_{F}}^{2}\tau /2$ is the diffusion constant, and $m^{*}
$ is the electron effective mass. If we replace $D$ by $1/2m^{*}$, the above
equation is quite similar to the Schr\"{o}denger equation for an electron.
However, an important difference is that $\psi $ is not a two-component
spinor, but a three-component vector in the three-dimensional Hilbert space
of the angular momentum $J$=1.

>From Eq.~(\ref{de2}), the spin flux can be expressed as
\[
{\bf F}=-D({\bf \nabla}_{{\bf r}}+im^{*}\alpha {\bf J\times y})\psi \ ,
\]
where {\bf y} is the unit vector along $y$ axis. The boundary conditions at
$z$=$\pm$$d/2$ are no spin flux across each boundary. Hence,
\begin{equation}\label{boundary}
D(-i\frac{\partial}{\partial z} +
m^*\alpha J_{x})\psi \,|_{z=\pm\frac{d}{2}} = 0\, .
\end{equation}

For an extended 2D electron gas with $d$$\rightarrow$$\infty$, it is easy
to see from Eq.~(\ref{de2}) that the relaxation rates of homogeneous spin
density are given by the eigenvalues of the operator
$Dm^{*2}\alpha^2(J_z^2$+$J_x^2)$. The two eigenvalues correspond to the
transverse (spin polarized in $xz$ plane) D'yakonov-Perel' spin relaxation
rate 1/$\tau_s$$\equiv$$Dm^{*2}\alpha^2$, and the longitudinal (spin
polarized in $y$ direction) D'yakonov-Perel' spin relaxation rate
2/$\tau_s$. In a 2D channel of electron gas, such homogeneous solution of
Eq.~(\ref{de2}) with spin polarized along the channel, which satisfies the
boundary conditions Eq.~(\ref{boundary}) also exists in the form of
$\psi_{\pm 1}$=$\pm$$1/\sqrt{2}$ and $\psi_0$=0. This state relaxes with the
rate 1/$\tau_s$ as in an extended 2D electron gas.

However, we will prove in this paper that in a narrow 2D channel there exist
inhomogeneous diffusion modes with much lower spin relaxation rates. To
derive these modes we will perform a canonical transformation
$\psi$=$U(z)\tilde{\psi}$, where $U(z)$=$\exp ($-$iJ_xz/L_s)$ with
$L_s$=$1/m^*\alpha$. The operator $U(z)$ transforms Eq.~(\ref{de2}) to
\begin{equation}\label{de3}
D\left[ -i\frac{\partial}{\partial x} - m^*\alpha \tilde{J}_z(z)
\right]^2\tilde{\psi} + D\frac{\partial^2}{\partial z^2}\tilde{\psi}
=\Gamma \tilde{\psi} \, ,
\end{equation}
with the boundary condition
$(\partial\tilde{\psi}/\partial z)|_{z=\pm d/2}$=0, where
$\tilde{J}_z(z)$=$U^{-1}(z)J_zU(z)$. If $d$$\ll$$L_s$, $U(z)$ can be
expanded in powers of the small parameter $z/L_s$. To the second order we
get $\tilde{J}_z(z)$=$J_z$+$J_yz/L_s$-$J_z(z/L_s)^2/2$ in which the last
two terms will be treated perturbatively. The lowest order perturbation
result gives the eigensolutions of Eq.~(\ref{de3})
\begin{equation}\label{eigen}
\tilde{\psi}_{M,k,m}(x,z)=\exp (ikx)\chi_m(z)\Psi_M \, ,
\end{equation}
where $M$ and $\Psi_M$ are eigensolutions of $J_z$ with $M$=$\pm$1 and 0.
The $\chi _{m}(z)$ functions are $\chi _{_{2n}}(z)$=$\cos (2\pi zn/d)$,
and $\chi_{_{2n+1}}(z)$=$\sin [\,\pi z(2n$+$1)/d\,]$ with integer $n$.

The unperturbed relaxation rates are readily obtained from Eq.~(\ref{de3})
as
\begin{equation}\label{rate}
\Gamma_{M,k,m}^0 = D(\pi m/d)^2 + D(k-ML_s^{-1})^2 \, .
\end{equation}
Hence, the modes with $m$$\neq$0 relax very fast. On the other hand, the
modes with $m$=0, which correspond to $\chi_{_0}(z)$=constant, can have
very low spin relaxation rates, which become zero for $k$=$ML_s^{-1}$.
However, corrections due to terms depending on $dL_s^{-1}$ in
Eq.~(\ref{de3}) make these rates finite. Using the standard perturbation
theory to the second order of $dL_s^{-1}$, we find
\begin{equation}\label{corr}
\Gamma_{M,k,0} = \Gamma_{M,k,0}^0 + \frac{(2-M^2)d^2}{24\tau_sL_s^2}
\end{equation}
for $|k$-$ML_s^{-1}|$$\ll$$L_s^{-1}$. The above equation indicates that
although the relaxation rates of the modes with $k$=$ML_s^{-1}$ become
finite, their relaxation times retain much longer than the D'yakonov-Perel'
relaxation time $\tau_s$.

There are three slowly relaxing diffusion modes with $m$=0, $M$=1, 0, -1,
and $k$=$ML_s^{-1}$. The two modes for $M=\pm 1$ correspond to inhomogeneous
spin distributions along the channel ($x$ axis) with wavevectors
$k$=$\mp$$L_s^{-1}$. Linear combinations of these modes give periodic spin
density distributions {\bf S}($x$) with electron spins rotating in the plane
perpendicular to $z$-axis. Hence, such a spin density has a finite
projection along the channel. The third mode for $M$=0 corresponds to
electron spins pointing along $z$-axis. The lowest relaxation rate is then
achieved by an inhomogeneous  spin density distribution along the channel.
We notice that although for $m=$0 the eigenfunctions given by
Eq.~(\ref{eigen}) are independent of $z$, their corresponding spin
distributions {\bf S}({\bf r}) depend on $z$. This is because
{\bf S}({\bf r}) is expressed via the $\psi$ function, which is obtained by
applying the $z$-dependent unitary transformation $U(z)$ to
$\tilde{\psi}_{M,k,0}(x)$.

We consider a stationary spin distribution in a channel of finite length
$L$$\gg$$d$, with an applied electric voltage $V$ between $x_l$=-$L$/2 at
the left and $x_r$=$L$/2 at the right. This voltage can produce a
magnetization in the channel, as was first proposed by Aronov~\cite{aronov}.
The channel is connected with two ferromagnetic contacts, and a spin flux
$F_l$ is injected into the channel at $x$=$x_l$. At $x$=$x_r$ a spin flux
$F_r$ is then collected. The corresponding boundary conditions are
\begin{equation}\label{boundary2}
D(-\frac{\partial}{\partial x} + im^*\alpha J_z)\psi |_{x=x_{l,r}}
= F_{l,r} \, .
\end{equation}
The stationary spin density in the channel can be derived from
Eq.~(\ref{de3}) with above boundary conditions. However, our goal is to
calculate the change of DC resistance associated with the spin transport,
which can be measured experimentally. To tackle this problem, we will use
the formalism developed by Johnson and Silsbee~\cite{johnson}. The
semiconductor-ferromagnet contacts are replaced by symmetric tunnel
junctions with identical conductances $G$=$Ae^2N(E_F)v_{_F}t/2$, where $A$
is the channel cross section area, $t$$\ll$1 the transmission probability,
and $N(E_F)$ the density of states at the Fermi energy.

A DC current through the system will cause interfacial voltage drops
$\Delta V_l$ across the left junction, and $\Delta V_r$ across the right
one. Gradients $\Delta S_{l,r}$ of nonequilibrium magnetization across the
tunnel junctions are also generated. Following Ref.~\onlinecite{johnson},
the spin fluxes through the junctions can be written as
\begin{equation}\label{flux}
F_{l,r} = -G\,[\,(\eta_{l,r}/2e)\Delta V_{l,r} -
(\xi /e^2)\Delta S_{l,r}\,] \, ,
\end{equation}
where parameters $\eta_l$ and $\eta_r$ with $|\eta_{l,r}|$$<$1 depend on
the magnetizations of the ferromagnetic contacts. $\eta_l$ and $\eta_r$
have same sign if the magnetic polarizations in contacts are parallel, and
opposite signs if antiparallel. The first term in Eq~(\ref{flux})
represents the spin injected by the applied voltage, and the second term
with $\xi$$\simeq$1 is the spin transport driven by gradients
$\Delta S_{l,r}$. If in ferromagnets this magnetization relaxes
sufficiently fast, $\Delta S_{l,r}$ are determined mainly by the spin
polarizations in the channel at $x_l$ and $x_r$, and so
$\Delta S_{l}$=-$\psi (x_l)$ and $\Delta S_{r}$=$\psi (x_r)$. Under
stationary condition, the DC current is
given by~\cite{johnson}
\begin{equation}\label{dc}
I = -G\,[\,\Delta V_{l,r} - (\eta_{l,r}/2e)\Delta S_{l,r}\,] \, .
\end{equation}

In this paper we will investigate the two cases that the magnetizations of
the ferromagnet contacts are either polarized along the $z$ axis
(perpendicular to the channel) or along the $x$ axis (parallel to the
channel). For each case, the magnetizations of the two contacts may be
parallel or antiparallel. In terms of the momentum operator {\bf J}, the
magnetization is represented by the $M$=0 state if it is along the $z$ axis,
but by a linear combination of the $M$=$\pm$1 states if it is along the $x$
axis. Hence, the corresponding injected fluxes are represented by
$F_{l,r}$=$f_{l,r}\psi_0$ and $F_{l,r}$=$f_{l,r}(\psi_1$-$\psi_{-1})$,
respectively. Since $L$$\gg$d, the dominating contribution to the
stationary spin distribution comes from the modes with low relaxation rates.
The corresponding exponents of the stationary solutions of Eq.~(\ref{de3})
can be derived from the equation $\Gamma_{M,k,0}$=0. In this way, from
Eq.~(\ref{corr}) we obtain $k$=$k_M$$\equiv$$ML_s^{-1}$$\pm$$il_M^{-1}$,
where $l_M^{-1}$=$\sqrt{(2-M^2)/24}d/L_s^2$. In order to satisfy the
boundary conditions Eq.~(\ref{boundary2}), we have to make a proper linear
combination of the $\exp{(x/l_M)}$ and $\exp{(-x/l_M)}$ solutions.
Furthermore, to demonstrate the spin dynamics along the channel, we average
Eq.~(\ref{boundary2}) over $z$. It is easily to see that the corrections
which are linear in $d/L_s$ vanish after the averaging. If we ignore the
higher order corrections, we can represent $\psi$ as a linear combination
of the zero order eigenfunctions Eq.~(\ref{eigen}) with $m$=0 and $k$=$k_M$.
The spin density distribution is thus represented by
\begin{equation}\label{psi}
\psi = \sum_M \psi_M \,e^{iMx/L_s}\,[A_M\cosh (x/l_M) +
B_M\sinh (x/l_M)] \, .
\end{equation}

>From Eq.~(\ref{boundary2}) we derive
$A_0$=$(f_l$-$f_r)L_s/[2D\cosh (L/2l_0)]$ and
$B_0$=--$(f_l$+$f_r)L_s/[2D\sinh (L/2l_0)]$ for the magnetization along $z$
axis. Similarly, for the magnetization along $x$ axis, the coefficients
$A_{\pm 1}$ and $B_{\pm 1}$ are obtained as
\begin{eqnarray}
A_M &=& \frac{ML_s}{2D\,\cosh \frac{L}{2l_M}}[f_l\,\exp (\frac{iML}{2L_s})
- f_r\,\exp (-\frac{iML}{2L_s})] \, , \nonumber \\
B_M &=& -\frac{ML_s}{2D\,\sinh \frac{L}{2l_M}}[f_l\,\exp (\frac{iML}{2L_s})
+ f_r\,\exp (-\frac{iML}{2L_s})] \, . \nonumber
\end{eqnarray}
It is important to point out that although the boundary conditions require
the spin polarization at $x$=$\pm$$L/2$ to be along the $x$ axis, within
the channel the spin polarization rotates in the $xy$ plane due to the
exponential factors $\exp{(\pm iMx/L_s)}$ in Eq.~(\ref{psi}). Since
$L_s$$\ll$$l_M$, the polarization makes many complete rotations within the
spin relaxation length $L_s$. As it will be shown below, this can lead to
an oscillation of the channel resistance.

Substituting the expressions of $A_M$ and $B_M$ into Eq.~(\ref{psi}), one
can find the spin densities at $x$=$\pm$$L/2$, and hence the nonequilibrium
magnetizations $\Delta S_{l,r}$. Together with Eqs.~(\ref{flux}) and
(\ref{dc}), we have a closed set of equations to solve numerically. However,
it is important to illustrate in analytical form how the resistance depend
on the spin transport through the 2D channel. For this purpose we consider
a sample of length $L$$\simeq$$l_M$, and having a low enough transmission
probability $t$ such that the parameter $\kappa$$\equiv$$tl_M/l$$\ll$1,
where $l$ is the electron mean free path. From Eqs.~(\ref{flux})-(\ref{psi}),
we readily derive the spin-transport correction to the DC resistance as
\begin{eqnarray}\label{res}
\Delta R &=& (2\kappa /G)(\eta_l^2 + \eta_r^2)\coth (L/l_M)
\nonumber \\
&& - (2\kappa /G)\,\eta_l\eta_r\,\cos (ML/L_s)/\sinh (L/l_M) \, .
\end{eqnarray}
The factor $\eta_l\eta_r$ in the second term at the right hand side is
positive if the magnetizations of the two ferromagnetic contacts are
parallel, but negative if antiparallel. Furthermore, when the magnetization
is along the channel, $M$=$\pm$1 and so this second term oscillates as a
function of $L/L_s$. The amplitude of oscillation decreases with
increasing $L/l_M$. However, due to the slowdown of the spin relaxation,
we can manipulate the sample parameters such that $l_M$$\simeq$$L$$>$$L_s$.
Then, our theoretically predicted resistance oscillation can be observed
experimentally with a gate to change the value of $L_s$.

In our analysis above we have neglected the Dresselhaus contribution to the
SOI, because in narrow gap systems which we are interested in, the Rashba
contribution dominates the SOI. In quantum wells the Dresselhaus
contribution contains a linear term and a cubic term in electron momentum.
It can be shown that the cubic term gives rise to an additional spin
relaxation which is independent of the channel width $d$. Hence this term
imposes a limit on the slowdown of spin relaxation. The corresponding
relaxation rate is estimated to be insignificant for narrow gap quantum
wells. However, in GaAs based quantum wells the cubic term is not
negligible~\cite{pp}, and can wash out the slowdown of spin relaxation.

Besides the classical spin diffusion, the spin relaxation slowdown also
affects the weak localization corrections to transport parameters. In
systems with strong enough SOI, the sign of such a correction to conductance
is determined by the competition between the triplet and the singlet
component of the Cooperon propagator~\cite{alt}. The lifetime of the triplet
is equal to the spin relaxation time. If the temperature is not very low, in
a sufficiently narrow channel~\cite{met} this time can increase and becomes
comparable to the dephasing time of the singlet. Hence, the sign of the
correction can change from positive to negative.

The weak localization corrections to the spin diffusion coefficient and the
D'yakonov-Perel' relaxation rate also contain contributions from the triplet
and the singlet components of Cooperon~\cite{us2,us}. The spin diffusion
coefficient is included in the factor $\kappa$ in Eq.~(\ref{res}).
Therefore, by measuring the change of channel resistance $\Delta R$ with a
controlled gate voltage or a weak magnetic field, the variations of spin
diffusion coefficient can be investigated experimentally.

We acknowledge the support of the Royal Swedish Academy of Science under the
Research Cooperation Program between Sweden and the former Soviet Union,
Grant No. 12527.

\widetext


\begin{references}
\bibitem{divi}  D. P. DiVincenzo and D. Loss, Superlattices and
Microstructures {\bf 23}, 419 (1998); B. E. Kane, Nature {\bf 393}, 133
(1998)

\bibitem{nitta}  J. Nitta, T. Akazaki and H. Takayanagi, Phys. Rev. Lett.
{\bf 78}, 1335 (1997).

\bibitem{datta}  S. Datta and B. Das, Appl. Phys. Lett. {\bf 56}, 665
(1990); S. Gardelis, C. G. Smith, C. H. W. Barnes, E. H. Linfield and D. A.
Ritchie, cond-mat/9902057 (1999).

\bibitem{dp}  M. I. D'yakonov and V. I. Perel, Sov. Phys. Solid State {\bf 13%
}, 3023 (1972) (Fiz. Tverd. Tela {\bf 13}. 3581, (1971)); Sov. Phys. JETP
{\bf 33}, 1053, (1971) [Zh. Eksp. Teor. Fiz. {\bf 60}, 1954 (1971)].

\bibitem{ilp}  S. V. Iordanskii, Yu. B. Lyanda-Geller and G. E. Pikus, JETP
Letters {\bf 60}, 206 (1994).

\bibitem{pp}  F. G. Pikus and G. E. Pikus, Phys. Rev. B {\bf 51}, 16928
(1995).

\bibitem{us2}  A. G. Mal'shukov, K. A. Chao, and M. Willander, Physica
Scripta {\bf T66}, 138 (1996).

\bibitem{dress}  G. Dresselhaus, Phys. Rev. {\bf 100}, 580 (1955)

\bibitem{rash}  Yu. L. Bychkov and E. I. Rashba, J. Phys. C {\bf 17}, 6093
(1984)

\bibitem{aronov}  A. G. Aronov, JETP Lett. {\bf 24}, 32, (1976) [Pis'ma Zh.
Eksp. Teor. Fiz. {\bf 24}, 37, (1976)]

\bibitem{johnson}  M. Johnson and R. H. Silsbee, Phys. Rev. B {\bf 35}, 4959
(1987); Phys. Rev. B {\bf 37}, 5312 (1988).

\bibitem{met}  We assume that the channel stays in the metallic regime,
hence it must be wide enough, so that $L<Nl$, where N is the number of
transmission eigenstates.

\bibitem{alt}  B. L. Altshuler and A. G. Aronov, in {\it Electron-Electron
interactions in Disordered Systems}, edited by A. L. Efros and M. Pollak
(North-Holland, Amsterdam, 1985).

\bibitem{us}  A. G. Mal'shukov, K. A. Chao, and M. Willander, Phys. Rev.
Lett. {\bf 76}, 3794 (1996).
\end{references}
\end{document}